# Vitrification and glass transition of water: insights from spin probe ESR

Shrivalli N. Bhat, Ajay Sharma and S.V.Bhat[*]

*Department of Physics, Indian Institute of Science, Bangalore-560012, India*

Three long standing problems related to the physics of water viz, the possibility of vitrifying bulk water by rapid quenching, its glass transition, and the supposed impossibility of obtaining supercooled water between 150 and 233 K, the so-called "No Man's Land" of its phase diagram, are studied using the highly sensitive technique of spin probe ESR. Our results suggest that water can indeed be vitrified by rapid quenching, it undergoes a glass transition at ~ 135 K, and the relaxation behavior studied using this method between 165 K and 233 K closely follows the predictions of the Adam-Gibbs model.

Water, unlike most other liquids, is a very poor glass former. Attempts to vitrify it in the bulk by direct rapid cooling have been unsuccessful so far, though a number of indirect methods have been used to obtain amorphous ice [1]. The first laboratory preparation of amorphous solid water (ASW) was carried out [2] by depositing water vapour at low pressure on a cold (~140 K) surface. Rapid cooling of small aerosolized water droplets on to a metal cryoplate [3] or in to cold liquids [4] also results in vitrification of water. The material produced using such processes is called hyperquenched glassy water (HGW) since high cooling rates are utilized. On compression of hexagonal ice a high density phase of amorphous phase (HDA) is obtained which on decompression and heating results in a low density amorphous phase (LDA) of water [5,6].

It has also been difficult to determine if the amorphous water undergoes a glass transition, and if so, at what temperature [7]. Differential scanning calorimetry ( DSC ) results on pre-annealed amorphous water indicated a glass transition around 136 K [8]. This conclusion was supported [9] by indentation of LDA at 143 K showing that it is a viscous liquid at that temperature. Thermally programmed desorption and isotope exchange experiments [10] provided evidence for molecular translational diffusion in

ASW between 136 K and the crystallization temperature of ~155 K giving further credence to the proposal of $T_g$ of ~ 136 K. However, Angell and co-workers noted [11,12] that the fact that the excess enthalpy of HGW is still unrelaxed at 150 K, the crystallization temperature to cubic ice phase $I_c$, means that the $T_g$ needs to be revised to 165 ± 5 K. This proposal and the calorimetric data on which it is based is controversial [9] in spite of a more recent study [13] on inorganic glasses which by comparison concludes that what happens in amorphous water at ~136 K is a 'shadow glass transition'.

Another noteworthy feature of the phase diagram of water is the occurrence of homogeneous nucleation around 233 K [14] while cooling, below which supercooled water cannot exist. Coming from the low temperature side, if the amorphous water is heated above its $T_g$, up to the crystallization temperature of ~150K it can exist in the form of supercooled, ultraviscous liquid. The temperature range from 150 K to 233 K, however, is excluded for the occurrence of supercooled water and is called the "No Man's Land" [15] in the water phase diagram.

In this work, we report on an attempt to vitrify bulk water by direct rapid cooling, which, as we believe, turns out to be the first successful way of doing so. We study this amorphous ice produced by spin probe electron spin resonance (ESR) [16], and demonstrate that around 135 K the amorphous ice transforms from a rigid material to one that executes fast dynamics. Comparing this result with those of similar experiments done on well known glass formers, we conclude that amorphous water indeed undergoes a glass transition around 135 K. We also find that the technique is able to give information on the relaxation behavior over the entire temperature range from about 165 K to 300 K, which qualitatively fits the existing theoretical models.

The most direct method of determining the $T_g$ i.e. DSC, is unable to give unambiguous signature of the glass transition in samples that experience a very small change in $C_p$ at $T_g$, such as partially crystalline amorphous polymers or water. $T_g$ in such systems can still be estimated [17,18] by using the highly sensitive, though indirect, technique of spin probe ESR. In this work, we first demonstrate the power of the technique by determining the $T_g$ of two well-known monomer glass formers viz. propylene glycol (PG) and glycerol and subsequently apply it to the determination of the Tg of water.

The amorphous water samples were prepared by directly exposing a capillary (dia ~ 100 microns) containing a small amount of triple distilled water doped with about 0.1% by weight of 4-hydroxy TEMPO (i.e. TEMPOL), (Aldrich) to liquid helium (4.2 K) *in situ* in the ESR low temperature cryostat. The liquid helium transfer tube was modified such that a burst of liquid helium hits the capillary cooling it to 4.2 K almost instantaneously leading to the formation of vitrified water. In figure 1(a), we present a few typical ESR signals showing the temperature dependence of the ESR spectra of the radical in the range 4.8 K to 300 K for fast quenched water. As explained later the spectra give unambiguous indication of the formation of vitrified water. For comparison, we also carried out ESR experiments on slowly cooled (~1 K / min) water doped with the same radical from room temperature down to 4.2 K. Except for the much slower cooling rate, other parameters of the experiment were kept the same as before. A few representative spectra are presented in figure 1(b). The narrow, three line hyperfine pattern signifying the fast tumbling of the radical (see below) continues from room temperature down to ~ 232 K, close to the homogeneous nucleation temperature $T_N$ ~ 231 K of water. At 230 K and below, down to 4.2 K a single broad signal from aggregated TEMPOL is observed. Note that the gross features of the spectrum continue unchanged across the nominal freezing temperature of 273 K. This is because pure water in micron-sized capillaries does not crystallize at this temperature but continues in the supercooled state down to $T_N$.

The drastic change in the spectral shape across $T_N$ is noteworthy. Below $T_N$ since water has crystallized, it no longer accommodates the radicals inside (we recall that crystallization is one of the most efficient methods of purification), which aggregate outside the ice crystals and now because the sample is no longer dilute gives a broad signal. In addition, fast quenching experiments were carried out on propylene glycol (spectra shown in fig 1c) and glycerol (spectra not shown) samples and the results validate the technique as described below.

The appropriate spin Hamiltonian for the analysis of the spectra shown in fig. 1 is given by $H = \beta \mathbf{B.g.S} + \mathbf{S.A.I}$, where **B** is the applied magnetic field, **S** and **I** are the spins of the electron (=1/2) and the nitrogen nucleus (=1) respectively, '**g**' and '**A**' are the Zeeman and the hyperfine interaction tensors respectively [16]. $\beta$ is the Bohr magneton. The observed sensitive temperature dependence of the lineshape and the splittings of the spectra arise from the effect of the rate and nature of the tumbling motion of the spin probe in the host matrix on the tensorial interactions.

Across the glass transition, when the temperature is increased, the viscosity of the sample decreases from about $10^{13}$ poise corresponding to a transition from the slow tumbling regime with $\tau_c \geq 10^{-6}$ s, to a fast tumbling regime with $\tau_c \leq 10^{-9}$ s. Below $T_g$ the nitroxide probe exhibits a powder pattern reflecting the full anisotropy of the Zeeman ('**g**') and the hyperfine ('**A**') interactions. When the frequency of the dynamics increases such that the jump frequency f is of the same order of magnitude as the anisotropy of the hf interaction. i.e. $\sim 10^8$ Hz, the anisotropy of the interactions gets averaged out and a spectrum with reduced splitting and increased symmetry in the lineshape is observed. This splitting corresponds to the non-vanishing isotropic value of the '**A**' tensor and is observed at a temperature higher than but correlated with $T_g$. The crossover is reflected in a sharp reduction in the separation between the two outermost components of the ESR spectrum, which corresponds to twice the value of the z-principal component of the nitrogen hyperfine tensor, $2A_{zz}$, from $\sim 70$ G to $\sim 30$ G.

We plot $2A_{zz}$ vs T for fast quenched water, propylene glycol and glycerol in figure 2. The behaviour is similar to that observed in a large number of synthetic polymers and monomers where a characteristic temperature $T_{50G}$ has been identified which corresponds to the temperature at which $2A_{zz} = 50$ Gauss. A near universal, semi empirical, correlation has been found [17] between $T_{50G}$ and $T_g$, which is shown in the inset to figure 2. Indeed, the temperature $T_{50G}$ corresponds to the 'high frequency' glass transition temperature reflecting as it does the temperature at which the relaxation time crosses the window of $\sim 10^{-8}$ s expected at a temperature higher than but correlated with $T_g$.

From figure 2, it is observed that $2A_{zz}$ undergoes a sharp transition in all the three samples, at 162, 242.5 and 259.5 K respectively in water, PG and glycerol. Corresponding to these $T_{50G}$'s we read out from the correlation plot (inset to fig. 2) the $T_g$'s as 132, 171 and 185 K respectively. We note that the values obtained for PG and glycerol are close to the literature values obtained by standard DSC results [19] (171 K and 191 ± 5 K respectively). We therefore expect the technique to provide a reliable value of $T_g$ for water as well, and conclude that the $T_g$ of water is close to 135 K and not 165 K.

The temperature dependence of the dynamics of supercooled water as reflected in the behaviour of viscosity and the relaxation time has also been the subject of a number of theoretical as well as experimental studies [7]. The central topic that has been debated is a possible, anomalous crossover from fragile to strong nature while going from moderately supercooled region to deeply supercooled region i.e. on approaching $T_g$ from above. The data on viscosity close to $T_g$ have been scarce and those for $T > \sim 240$ K have been fitted to a model based on the Adam-Gibbs equation $\tau = \tau_0 \exp(A / T S_{conf})$ where $\tau$ is a characteristic relaxation time (e.g. $\propto$ the viscosity), $\tau_0$ and A are temperature independent constants and $S_{conf}$ is the molar configurational entropy. A number of studies have shown that the orientational relaxation time of the spin probe reflects the diffusive behaviour of the host though depending upon the system it may be either fully or partially coupled to the host dynamics. Using the spin probe ESR spectra, in the fast

motion regime, i.e. for T > $T_{50G}$ the τc's of the probe can be calculated [20] from the equation

$$\tau_c = [\{(h_{+1}/h_{-1})^{0.5} - 1\}\Delta H_{+1}] / (1.2 \times 10^{10}) \text{ s},  \tag{1}$$

where, as indicated on one of the signals in fig. 1(c), $h_{+1}$ is the peak to peak amplitude of the low field signal, $h_{-1}$ is that of the high field signal and $\Delta H_{+1}$ is the peak to peak width of the low field signal. In fig. 3, we plot the correlation time $\tau_c$ obtained for water and PG calculated using equation 1 vs $T_g/T$. Similar analysis was not attempted for glycerol because of lack of enough data points above $T_{50G}$. The data for PG fits a straight line Arrhenius dependence excellently whereas the anomalous nature of water is clearly seen. We compare this experimental result with the behaviour of the viscosity η of water obtained from ref. [21]. The η → T plot in this work covers a few experimental points in the limited temperature range (of moderately super-cooled region) and a theoretical curve based on the Adam – Gibbs model in a broader temperature range of $T_g/T$ varying from about 0.4 to 0.9. The applicability of the model has been corroborated by recent computer simulations [22]. Noteworthy is the close similarity between the $\tau_c$ behaviour obtained by us for almost the entire range and the viscosity curve. A perfect match is not expected between $\tau_c$ of the spin probe and the viscosity η of the host in view of the possibility of decoupling of diffusion and viscosity on approaching the glass transition [23]. However, the observed similarity of the qualitative features of the curves over the wide temperature range indicates that the probe relaxation is intimately connected with the host dynamics.

The $\tau_c$ data of supercooled water reported by us in fig. 3 is seen to cover a wide temperature range within the so-called "No Man's Land" where we observe narrow ESR signals signifying the presence of supercooled water. The key to the understanding this apparently puzzling result is in the structural and dynamic heterogeneity of water discussed over a long time [24-26]. Experimental proof of this is provided by the femtosecond mid-ir pump-probe spectroscopic evidence [27] for the occurrence of two types of relaxation: one very slow for strongly hydrogen bonded water molecules and the other fast for weakly H-bonded molecules indicating that two distinct molecular species exist in liquid water. Raman spectroscopic study of Nishi et al [28] gives evidence for four different types of water molecules: (1) "free" water, whose molecules do not participate in any hydrogen bonding, (2) weakly associated water, the molecules of which have one hydrogen bonded to a neighbouring water but the other one is not bonded, (3) strongly bonded water with molecules both the hydrogens of which are bonded and (4) icy water with tetrahedral coordination. It stands to reason that more weakly bonded water molecules (i.e. those that are less likely to have the tetrahedral structure) are easier to vitrify since one of the main reasons for water being a poor glass former is understood to be the fact that the symmetry of the ice crystals is consistent with that of short range tetrahedral bond ordering [29] present in liquid water. We believe that it is this "free" or/and free + weakly associated water, which forms a small fraction of the entire sample, that we are able to vitrify and study the dynamics of in the so-called "No Man's Land". This would explain why other techniques, not as sensitive as spin probe ESR, are not able to detect it. This model of co-existence of liquid water with crystalline (cubic) ice in the temperature range 140 – 210 K is also supported by the very recent results Souda[30] which are consistent with the earlier observation Jenniskens et al.[31] that viscous water persists in the temperature range along with cubic crystalline ice.

In summary, we have carried out spin probe ESR experiments on rapid quench-formed amorphous water and two well-known glass formers, propylene glycol and glycerol. In the latter two materials we show that the $T_g$'s read out from the correlation diagram between $T_{50G}$ and $T_g$ are consistent with the reported $T_g$'s obtained from DSC measurements. Carrying out a similar analysis of the spin probe ESR results on glassy water, we confirm that water undergoes a glass transition around ~135 K. The temperature dependence of the reorientational correlation time of the spin probe obtained over a wide temperature range is shown to be qualitatively similar to that predicted by the Adam-Gibbs model.

Acknowledgements: The authors gratefully acknowledge extensive and useful discussions with C.A.Angell, P. G. Debenedetti, D. Leporini and F.C Starr. G.P. Johari is thanked for helpful, critical comments and for bringing the latest work to the authors' attention.

*To whom correspondence should be addressed. E-mail: svbhat@physics.iisc.ernet.in

FIGURE LEGENDS

Fig. 1: ESR spectra of TEMPOL in (a) fast quenched and (b) slowly cooled water and (c) fast quenched propylene glycol (PG) at a few representative temperatures. For the fast quenched samples, (a) and (c), the radicals are trapped in glassy matrices at low temperatures and give rise to broad 'powder patterns' indicative of anisotropic '**A**' and '**g**' interactions. For the slowly cooled sample, (b), below the homogeneous nucleation temperature ~232 K, a single broad signal from aggregated TEMPOL, which is thrown out of the crystalline water is observed. The method of estimating the extrema separation $2A_{zz}$ and the amplitudes and the linewidths used in equation 1 for calculating $\tau_c$ is also indicated.

Fig. 2: Extrema separation ($2A_{zz}$) vs temperature for TEMPOL doped water (circles), propylene glycol (hollow squares) and glycerol (stars). All the three samples show sharp transitions of $2A_{zz}$. $T_{50G}$, the temperatures at which $2 A_{zz} = 50$ Gauss are indicated. The inset shows Correlation plot for $T_{50G}$ *vs* $T_g$ adapted from ref. 22. The line and the numbered hollow circles are from the work of Kumler (ref. [17]). The data points correspond to different polymer samples: 1: PDMS; 2: 0.42S/B; 3: 0. 62 S/B; 4: PS-706; 5: Polycarbonate. Ref. [17] has many more data points which have been left out for clarity. Our results are shown in the same graph for glycerol (star), propylene glycol (hollow square) and water (filled square). Not only the $T_{50G}$ for glycerol and PG lie exactly on the graph but the respective $T_g$ 's estimated match with literature values. The value of $T_g$ for water is estimated to be 132 K from the graph.

Fig. 3. The temperature dependence of the product $\tau_c$ x T for propylene glycol(PG) (hollow squares) and water (filled squares) plotted as log ($\tau_c$ x T ) vs $T_g$ / T. The line through the data of PG is a least square fit to a straight line whereas the smooth line through the data of water is a guide to the eye. In the same figure log η vs $T_g$ / T for water obtained from ref. 21 is also shown for comparison. The latter shows a few experimental points in a limited temperature range plotted over the theoretical curve determined assuming the validity of the Adam-Gibbs equation. It is seen that while the correlation time of propylene glycol follows the Arrhenius behavior excellently, $\tau_c$ of water shows qualitatively the anomalous behavior predicted by the theory.

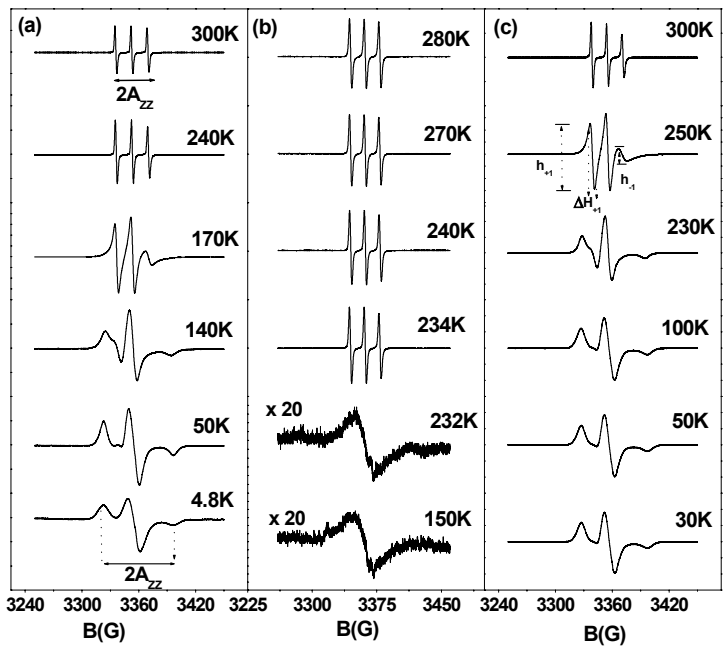

Figure 1: Bhat et al.,

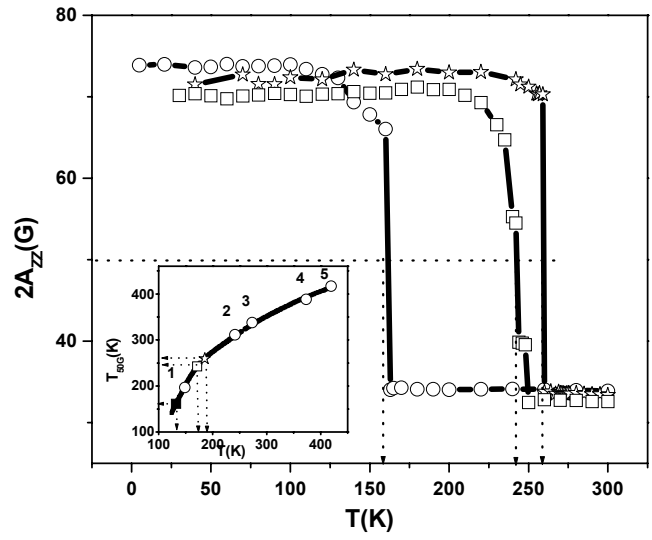

Figure 2: Bhat et al.,

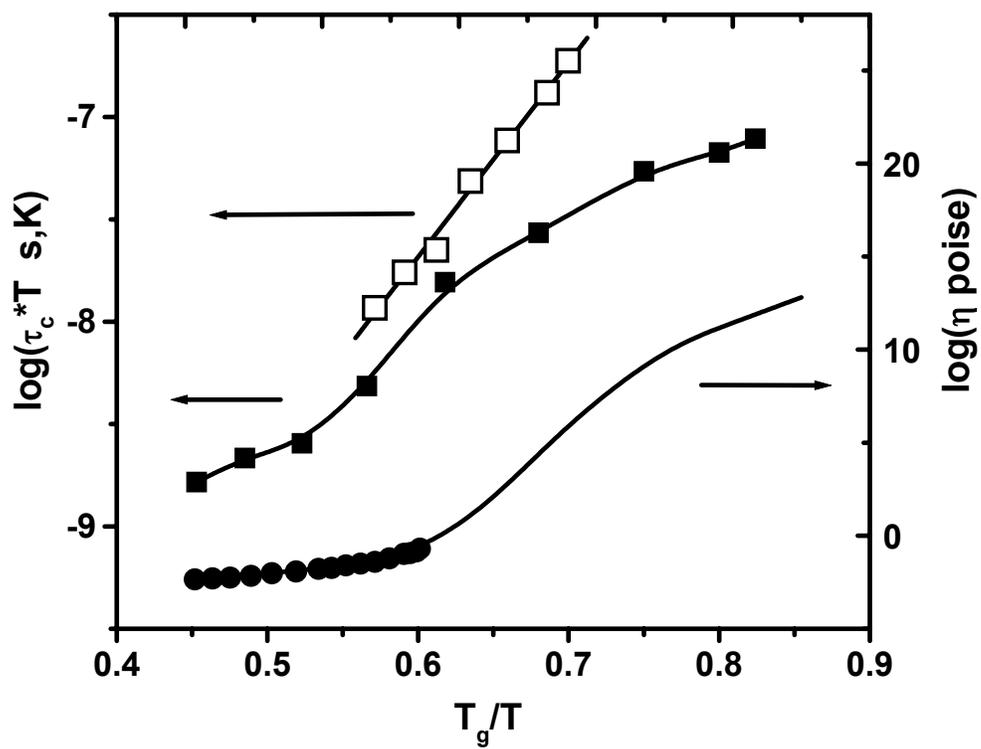

Figure 3: Bhat et al.,